\documentclass[11pt,pre,onecolumn,tightenlines,longbibliography,notitlepage]{revtex4-1}
\pdfoutput=1

\usepackage{xcolor}
\usepackage{verbatim}
\usepackage{graphicx}
\usepackage{siunitx} 

\newcommand{\vrat}{$\eta_{\mathrm{in}}/\eta_{\mathrm{out}}$}

\newcommand{\LonLc}{$L_{\mathrm{onset}}/\lambda_c$}

\begin{document}

\title{Delayed onset and the transition to late time growth in viscous fingering}

\author{Thomas E. Videb\ae k}
\affiliation{Department of Physics, The University of Chicago, Chicago, IL 60637, USA}

\begin{abstract}

Viscous fingering patterns form in confined geometries at the interface between two fluids as the lower-viscosity fluid displaces the one with higher viscosity. Previous studies have examined the most unstable wavelength of the patterns that form using both linear-stability analysis and the dynamics of finger growth in  the nonlinear regime.  Interesting differences in dynamics have been seen between rectilinear and radial geometries as well as between fluid pairs that are immiscible (with interfacial tension) or miscible (with negligible interfacial tension). This paper reports measurements of how all of these systems transition from the linearly unstable regime to their late time, nonlinear dynamics. In all four cases there is a region of stable or slow growth characterized by an onset length scale before fingers enter the late-time regime. For immiscible fluids in a rectilinear geometry this onset length is consistent with linear-stability analyses. All other cases are not adequately described by existing theory. In radial geometries, the onset length predicted from theory is an order of magnitude smaller than what is experimentally observed and has the incorrect scaling with dimensionless numbers. For miscible fluids in a rectilinear geometry the onset length is related to the development of steady-state structures within the confining dimension and cannot be explained by quasi-two dimensional theories. By combining the onset length with the finger growth rate measured after onset, the global patterns that form well into the late-time dynamics can be predicted.

\end{abstract}

\maketitle

\section{Introduction} %

Dynamic instabilities are the source for the inception and evolution of many pattern formation problems in hydrodynamic systems, many of which involve the motion of free interfaces. A common mode of growth in such physical systems results in the formation of long, branching structures, as seen in diffusion-limited aggregation~\cite{Witten1981,Sander85,Sander86}, directed solidification~\cite{langer1980instabilities}, river-network formation~\cite{tarboton1988fractal}, and many others. A prototypical system for studying branching growth has been the viscous fingering instability~\cite{Bensimon86,Homsy87,Couder00}, which occurs when a lower viscosity fluid displaces a higher viscosity one in a confined geometry. A typical experimental system for studying viscous fingering is the Hele-Shaw cell, which consists of two parallel flat plates whose separation, $b$, is much smaller than their lateral extent.

Previous studies of viscous fingering have calculated the expected width of fingers, $\lambda_c$, in the context of linear stability, which occurs in the early stages of growth~\cite{saffman1958penetration,paterson1981radial,paterson1985fingering,kim2009viscous,gadelha2009effects,nagel2013new,dias2013wavelength}. Excellent agreement between theory and experiment has been found in this regard. Outside the scope of linear stability, many studies have investigated the nonlinear regime of the patterns~\cite{Cheng08,bischofberger2014fingering,bischofberger2015island} and have characterized aspects of the large scale structures including the fractal dimension of the interface~\cite{Paterson84DLA,fernandez1990diffusion,Swinney05} and the growth rate of fingers~\cite{tan1988simulation,Chen89,booth2010growth,malhotra2015experimental}. 

\begin{figure*}
    \centering
    \includegraphics[width=0.6\linewidth]{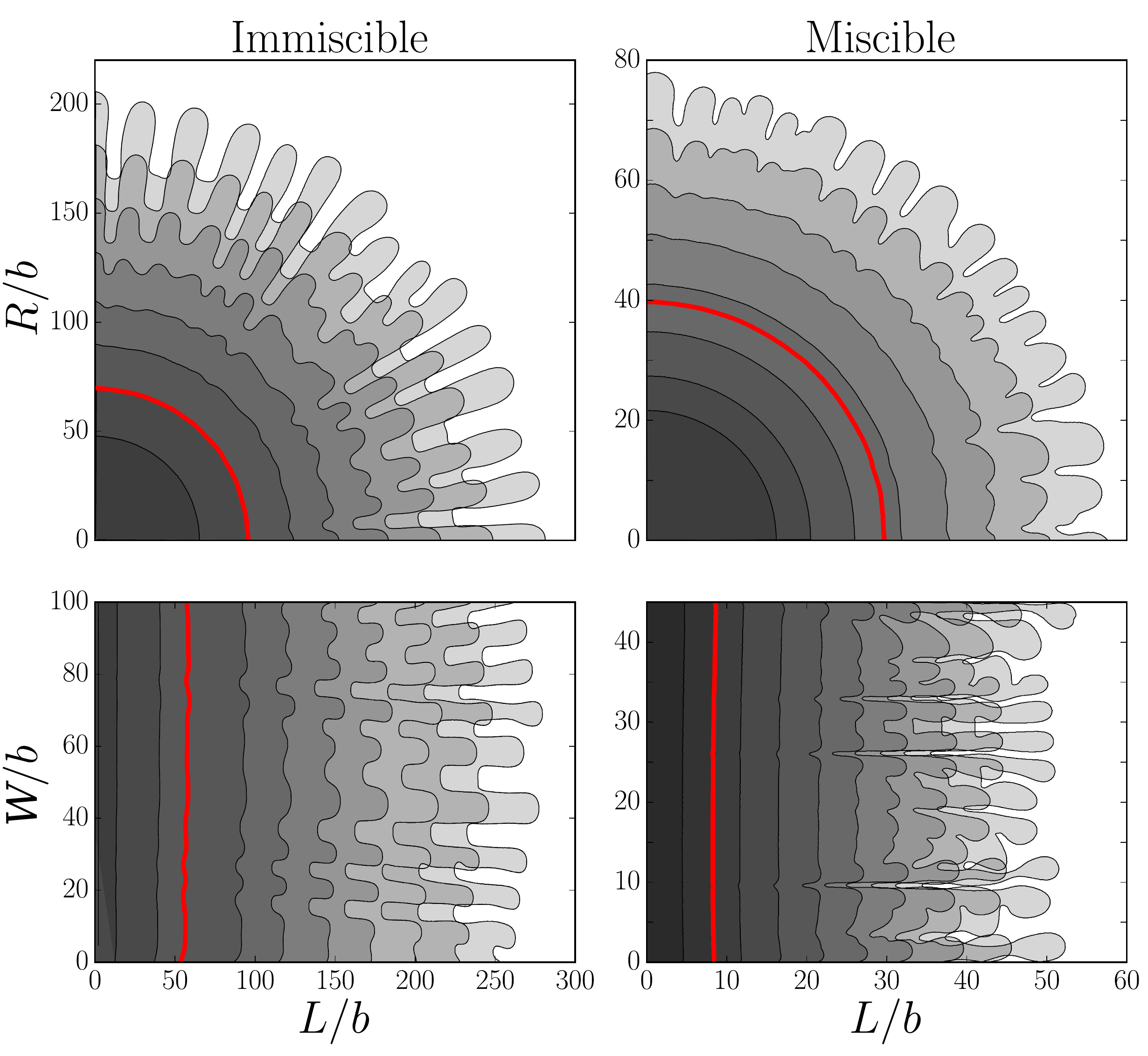}
    \caption{Contours at different times for radial and rectilinear geometries with both miscible and immiscible fluids. The red curve is the interface where fingering is first detected. $R/b$ and $L/b$ are radial and linear lengths rescaled by the gap dimension, $b$. $W/b$ is the distance along the width of the interface for the rectilinear geometry. Starting with the top left and going clockwise the fluid properties for these experiments are: $\eta_{\mathrm{in}}/\eta_{\mathrm{out}} = [0.35, 0.13, 0.10, 0.59]$ and $\Delta\eta=[460, 587, 350, 534]$ cP, with $b=[205, 485, 660, 660]$ \si{\micro}m, respectively.}
    \label{fig:OnsetQual}
\end{figure*}

However, one feature of the pattern dynamics has been underappreciated: there is a substantial delay after injection starts before the instability occurs and fingers start to grow. In the top panel of Fig.~\ref{fig:OnsetQual} the delayed onset can be seen in a radial geometry. The contours in this figure show the interface of an experiment at different moments in time. At the latest time (and therefore the furthest extent), fingers are fully developed. Looking at the earlier (inner) contours, one can follow how these fingers are formed; going to earlier times, the fingers get shorter and shorter until they eventually become immeasurably small. This point is denoted by the red curve; this does not occur at the injection point, but rather at a finite distance from the orifice. The existence of such a length scale, which I call the onset length, can be seen in the other panels of Fig.~\ref{fig:OnsetQual} for both radial and rectilinear geometries as well as immiscible and miscible fluids. This onset length is a robust feature of the instability. 

Considering the success of existing theory at determining the wavelength for the fingers that form, one might expect that the onset length should also fall within the scope of existing linear-stability~\cite{paterson1981radial,kim2009viscous,nagel2013new}.  However, this paper will show that this is not the case by carefully investigating the features of delayed onset in rectilinear and radial geometries for both immiscible and miscible fluids. All four of these systems show the same qualitative behavior, there is a length scale up to which the patterns must reach before fingers become readily apparent and begin growing rapidly. Despite the similarities of the delay in all four of these systems, the linear immiscible case behaves in a different way: even though it has an apparent delay in the growth dynamics, this system does not have a length scale associated with a region of true stability. This distinguishes it from the other systems. Even though it has been noted that a radial geometry should provide a period of stable growth~\cite{paterson1981radial,kim2009viscous,nagel2013new}, I will show that expectations from existing work are much too small to account for the stability found in experiments and have the wrong dependence on system parameters.

When the two fluids are miscible, the character of the viscous fingering instability is also greatly altered. The most dramatic aspect of this is an increased stability of the system so that when the ratio of the inner fluid viscosity, $\eta_{\mathrm{in}}$, to the outer fluid viscosity, $\eta_{\mathrm{out}}$, is in the range \vrat$>2/3$ fingering has not been observed~\cite{Lajeunesse97,lajeunesse1999miscible,bischofberger2014fingering}. 
This stability coincides with a change in the gap structure for the inner fluid. Since there is no surface tension, capillarity does not force the inner fluid to fully fill the gap, allowing for additional structure. The connection of the gap structure to the presence of the instability has not been fully understood. Here I will demonstrate how some aspects of viscous fingering that differ between miscible and immiscible fluids can be brought into a common framework by accounting for the changing gap structure.

Recent work has highlighted that there can be a substantial inner displaced region in the fingering patterns~\cite{bischofberger2014fingering,bischofberger2015island} that influences the overall character of the structures that form in the nonlinear regime. Previous work measured the ratio of the length of fingers to the radius of the fully displaced region at the interior of the pattern. The experiments demonstrated a delay in the instability onset with a cross-over time between fast and slow growth for miscible fluids in a radial geometry. There is a strong connection between the onset length found here and the characteristic timescale reported in the earlier work. So far, the behavior of the size ratio is unexplained. By combining measurements of the onset length with the growth rate of fingers, the dynamics of the size ratio is captured quite well. It demonstrates a connection between the delayed onset and the late time patterns that form.

Due to the complexities that are introduced by using different geometries and by having miscible versus immiscible pairs of fluids, the particulars of each system will be discussed separately. Section II covers the methods used in the experiments and simulations. Section III demonstrates the general trends of how the delayed onset depends on different fluid parameters and emphasizes what is common between the different systems. Section IV is a discussion of the onset length; it is divided into subsections for each geometry and type of fluid. Section V shows how the onset length and finger growth rate can be coupled to explain the size ratio. Finally, Section VI is a discussion of these results and their implications.

\section{Methods}

This paper reports experiments conducted in both radial and rectilinear Hele-Shaw cells. These cells consist of two large, flat glass plates of 1.9 cm thickness with a uniform gap, $b$, between them. The radial plates have a 14 cm radius and the rectilinear plates are 17.8 cm wide by 30.5 cm long. The gap spacing (constant for each experiment) is varied between 75 \si{\micro}m and 1145 \si{\micro}m using spacers placed at the perimeter of the cells. Details about these setups are described in the Methods section of ref.~\cite{videbaek2019diffusion}. 

Both miscible and immiscible experiments use pairs of water-glycerol mixtures, silicone oils, and mineral oils. The injection rate is controlled by a syringe pump (NE-1000 from New Era Pump Systems Inc.); the injection rate is varied in different experiments between 0.2 mL/min to 70 mL/min. For the miscible experiments, the injection is fast enough such that the P\'eclet number, which is the dimensionless ratio of the effects of advection to the effects of diffusion, is large enough to ensure that diffusion is not influencing the dynamics. To measure the fraction of the gap that the inner fluid occupies, the fluids are dyed and the measured intensity is compared to a calibrated cell of known thickness. Aqueous fluids are dyed with Brilliant Blue G at a concentration of 0.4 mg/mL, and oils are dyed with Oil Red O with a concentration near the saturation point (Alfa Aesar). Fluid viscosities are measured using the SVM 3001 viscometer and MCR 301 rheometer (Anton Paar). 

\begin{figure*}[t]
    \centering
    \includegraphics[width=\linewidth]{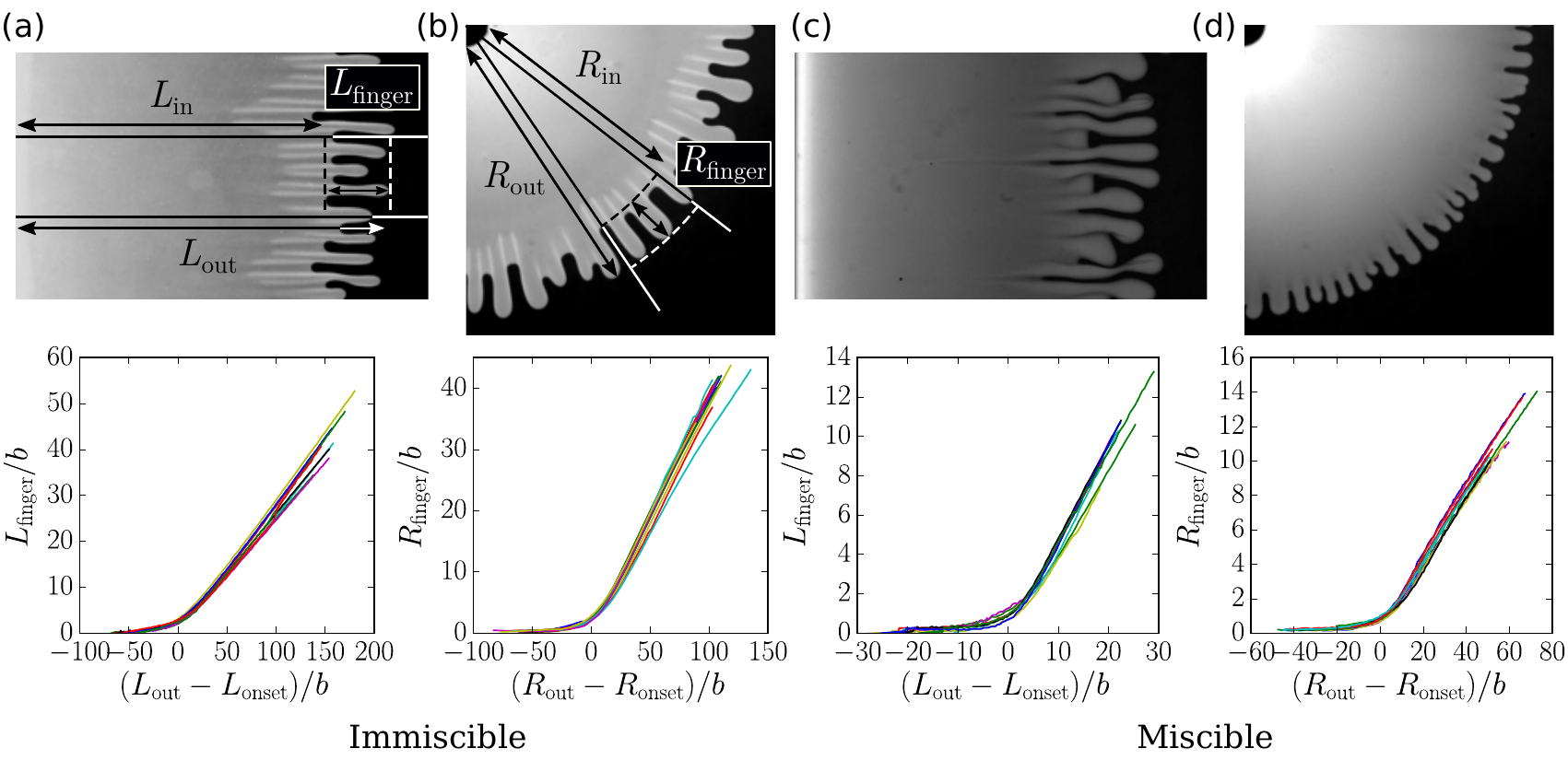}
    \caption{(a) and (b) show measurements of the finger length $L_\text{finger}$ and $R_\text{finger}$ against the pattern size, $L_\text{out}$ or $R_\text{out}$, for immiscible fluids in rectilinear and radial geometries respectively. The white lines just behind the fingers are artifacts from imaging due to the index of refraction difference between the two fluids and a slight curvature to the interface between them. Note that $L_\text{out}$ and $R_\text{out}$ has been shifted by the transition point $L_\text{onset}$ and $R_\text{onset}$. All lengths are non-dimensionalized by the plate spacing, $b$. (c) and (d) show the same measurements for miscible fluids.}
    \label{fig:Measure}
\end{figure*}

To quantify the behavior of the onset, I focus primarily on the dynamics of the finger length. 
Note that $L$ and $R$ are used to reference lengths in the rectilinear and radial geometries respectively. The length of a finger, $L_\mathrm{finger}$ for the rectilinear cell and $R_\mathrm{finger}$ for the radial cell, is defined as the difference between the furthest extent of the inner fluid from the injection point, $L_\mathrm{out}$ or $R_\mathrm{out}$, and the closest point to the inlet where the inner fluid fills the gap, $L_\mathrm{in}$ or $R_\mathrm{in}$, see Fig.~\ref{fig:Measure}a and b.  
In the preparation of the initial interface for the rectilinear geometry or the determination of the center of the pattern for the radial one, there is always some initial perturbation along the interface that can be mistaken for a very small but measurable finger length at the start of the experiment. Since the purpose of this paper is to look in detail at the earliest stages of \textit{growth}, these perturbations are subtracted off and the finger is defined as the difference from this initial state.

Numerical simulations of the flow within the gap for miscible fluids were performed using COMSOL. To model the inner and outer fluids, the \textit{Transport of Concentrated Species} module was used to track the mass fraction, $c$, of the inner fluid. Advection of the mass fraction is coupled to Stokes flow, $\eta\nabla^2\mathbf{u}=\rho\nabla p$, with diffusion modeled by a Fick's law: $\partial_t c = D\nabla^2 c$, where $\eta$ is viscosity, $\textbf{u}$ is the velocity field, $\rho$ is density, $p$ is the pressure, and $D$ is a diffusion coefficient. The simulation domain has a height of $b$ and a length of $L$. The flow satisfies no-slip conditions at the top and bottom of the simulation domain, has an average velocity, $U$, on the inlet side and a condition of constant pressure on the outlet side. The viscosity is related to mass fraction by
$\eta(c) = \eta_{\mathrm{out}}\exp(c \ln(\eta_{\mathrm{in}}/\eta_{\mathrm{out}}))$, with $c=1$ corresponding to the inner fluid. The parameters used are $D=10^{-12}\ \mathrm{m^2/s}$, $\eta_{\mathrm{in}}=1\  \mathrm{cP}$, $U=0.001$ m/s, $\rho=1\ \mathrm{g/cm^3}$, $b=1$ mm, and $L$ ranging from 15 mm to 100 mm; $\rho$ is the same for both fluids. The domain is discretized on a square mesh with side length 0.01mm. The value of $\eta_{\mathrm{out}}$ is varied to achieve a desired viscosity ratio.

\section{Linear growth transition}
\label{sec:lineartransition}

The images in Fig.~\ref{fig:OnsetQual} show qualitatively similar behavior for both radial and rectilinear cells as well as for miscible and immiscible pairs of fluids. When the pattern size is small (close to the inlet) the interface looks smooth with no visible fingers. At later times, fingers are clearly visible and located all along the interface. At an intermediate time, the system passes through a point where fingers just become detectable. Extrapolating a finger's length back to the point where it disappears provides a measure of the onset length, $L_\text{onset}$ or $R_\text{onset}$. The onset point is measured locally, as seen in the images for immiscible fluids in Fig.~\ref{fig:Measure}a and b, either for a small section of interface for the rectilinear cell or for a small angular section for the radial cell. An average is taken over all sections of the interface that are free of initial defects (these could include air bubbles or pinning points on the plate surfaces). The plots in Fig.~\ref{fig:Measure} show the behavior of $L_\text{finger}$ and $R_\text{finger}$ with respect to the pattern size, $L_\text{out}$ or $R_\text{out}$. At larger pattern size, both radial and rectilinear geometries show a regime where the growth of the finger length with pattern size is constant. For the rectilinear case, I fit this regime to a first-order polynomial and extract both the onset length $L_\text{onset}$ (where the linear fit extrapolates to zero finger length) and the growth rate of the patterns, $\Lambda_l\equiv dL_\text{finger}/dL_\text{out}$. For the radial case $R_\text{onset}$ and $\Lambda_l\equiv dR_\text{finger}/dR_\text{out}$ are similarly acquired.

It is important to make a distinction between a true onset transition, where the pattern first starts to become unstable, and a crossover to late-time growth where the finger evolution is no-longer governed by linear-stability analysis. Thus, the extrapolated length discussed by Fig.~\ref{fig:Measure} does not mark a transition from stability to instability but is a transition to the late-time growth regime which is qualitatively similar in all four cases.  However, by looking more carefully at the early-time growth, one can identify a true transition  between stability and instability for three of the four cases.  

\begin{figure*}[t]
    \centering
    \includegraphics[width=\linewidth]{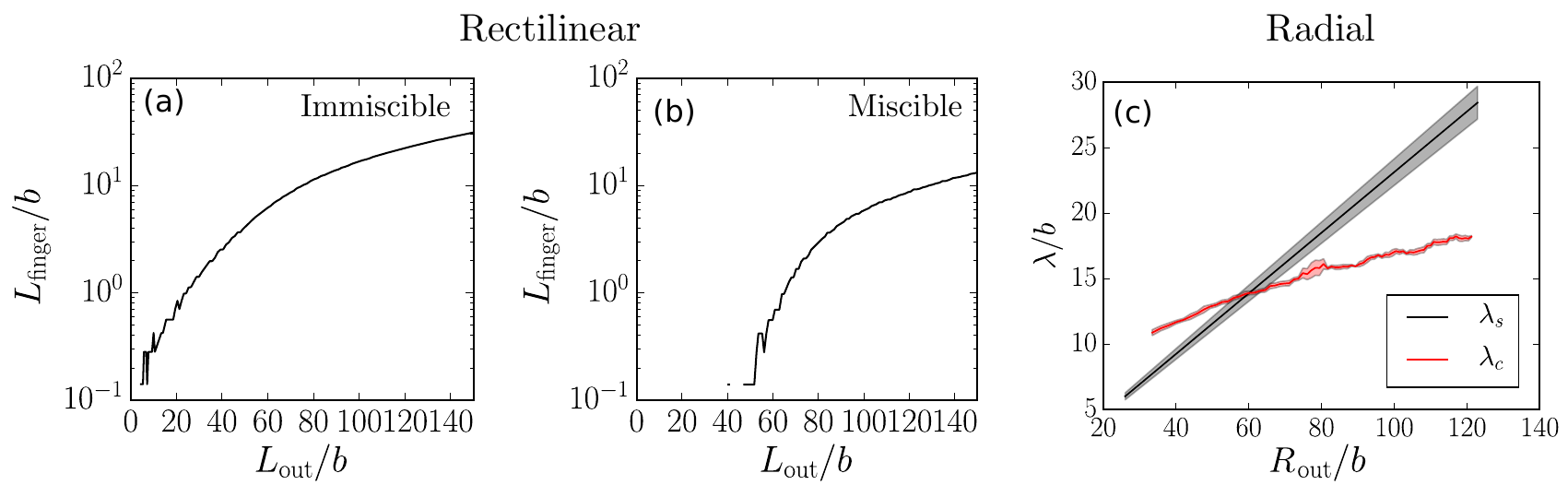}
    \caption{(a) and (b) show the growth of $L_\mathrm{finger}$ at early times in a rectilinear geometry for the immiscible or miscible case respectively. Note that the finger size is shown on a logarithmic scale. (c) For the radial immiscible case the comparison of the arc length of a disturbance, $\lambda_s$, and the allowed most unstable wavelength, $\lambda_c$, based on the local interface velocity.}
    \label{fig:EarlyDelay}
\end{figure*}

Only the rectilinear immiscible case does not show this delayed onset as can be seen in Fig.~\ref{fig:EarlyDelay}a which shows the growth of $L_\mathrm{finger}$ at early times. It shows that growth occurs as soon as the dynamics begin. In contrast, Fig.~\ref{fig:EarlyDelay}b shows the early stages of growth for the rectilinear miscible case, where there is a substantial delay before fingers begin to appear. In the radial case note that the existence of a most unstable wavelength, $\lambda_c$, and the appearance of a set number of fingers gives a geometrical constraint on the circumference at which the instability can first form. Figure~\ref{fig:EarlyDelay}c shows the expected $\lambda_c$ and the arc length of the angular width of the fingers, $\lambda_s$.   There is a single point at which they cross. I argue below that this delay length, seen in all but the rectilinear immiscible case, is associated with a stable-unstable transition.

\begin{figure*}
    \centering
    \includegraphics[width=.8\linewidth]{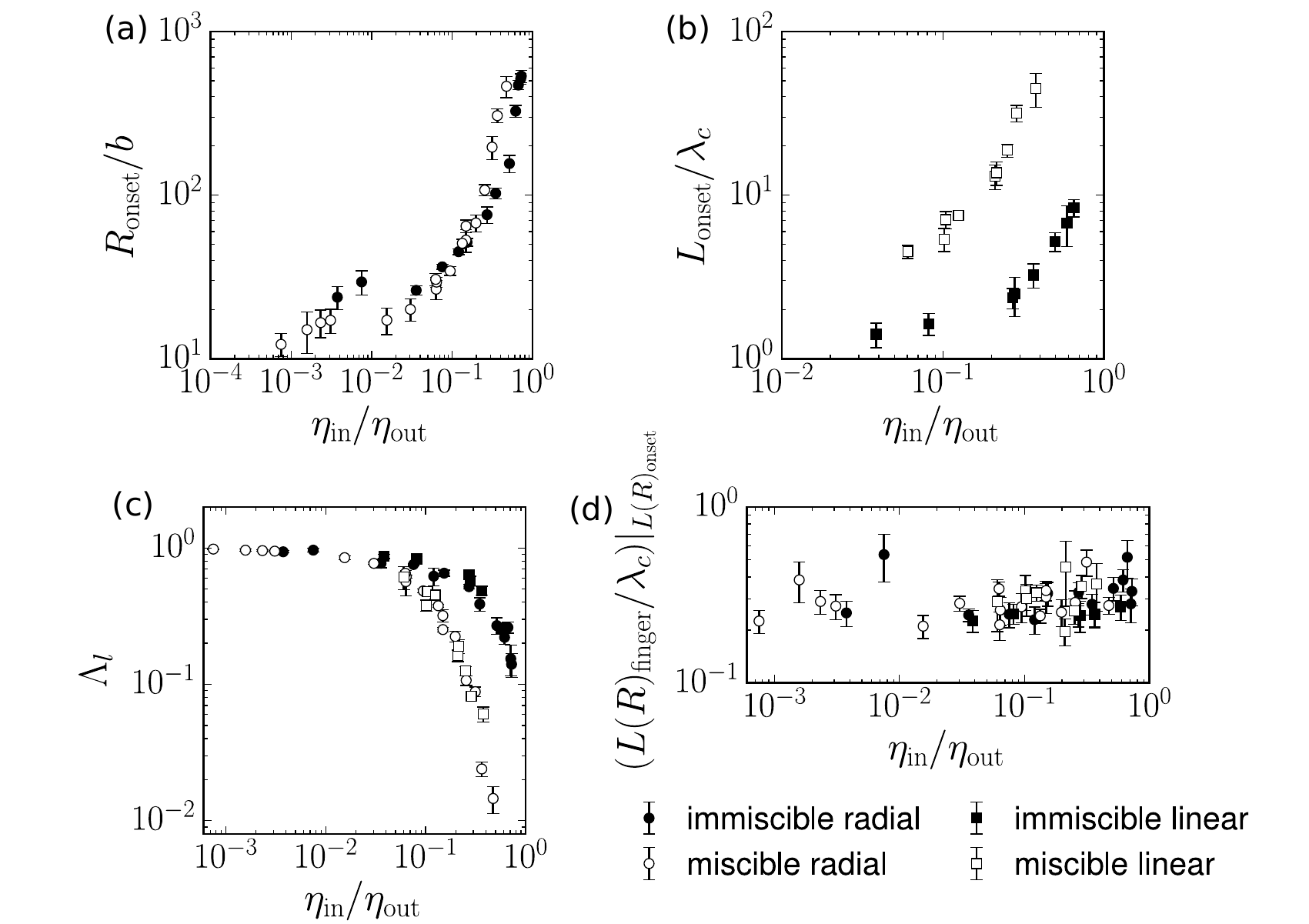}
    \caption{Onset length versus viscosity ratio, \vrat. For the radial geometry (a) $R_\text{onset}/b$ is shown while in the rectilinear geometry (b) $L_\text{onset}/\lambda_c$ is shown. (c) shows the growth rate, $\Lambda_l$, versus \vrat\ for all four cases. $\lambda_c$ used for rescaling has been measured in experiment; for the immisicble case $\lambda_c$ matches theory well~\cite{saffman1958penetration,paterson1981radial} and for the miscible case I find $\lambda_c/b=2.97\pm0.33$. (d) shows the measured finger length, $L_{\mathrm{finger}}$ or $R_{\mathrm{finger}}$, at the transition point. All systems show a finger length of $(0.30\pm0.07)\lambda_c$ at the transition.}
    \label{fig:OnsetData}
\end{figure*}

Figures~\ref{fig:OnsetData}a and b show how $R_\text{onset}$ and $L_\text{onset}$ depend upon the viscosity ratio for both the radial and rectilinear geometries respectively. $R_\text{onset}$ is non-dimensionalized with $b$ while $L_\text{onset}$ is non-dimensionalized with $\lambda_c$. The choice of $b$ or $\lambda_c$ is justified in the next section, which looks at the scaling of $L_\text{onset}$ and $R_\text{onset}$ with the capillary number. In all four cases there is a strong increase in the value of the onset length as $\eta_\text{in}/\eta_\text{out}\rightarrow 1$. The radial geometry shows rather similar values of $R_\text{onset}/b$ for the miscible and immiscible pairs of fluids; however, in the rectilinear case the miscible system has larger onset length compared to the immiscible one. The strong dependence upon viscosity ratio is also seen in the growth rate, $\Lambda_l$, shown in Fig.~\ref{fig:OnsetData}c. Lastly, Fig.~\ref{fig:OnsetData}d shows the size of fingers that are observed at the transition point. Note that this data has been rescaled by $\lambda_c$ rather than by $b$. In all cases this transition occurs when the finger length becomes $(0.30\pm0.07)\lambda_c$, denoting the size of perturbations at which the regime of linear stability ends.

Note that the behavior of the miscible fluids is slightly different from the immiscible ones as the viscosity ratio nears unity.   In the miscible fluids, $\Lambda_l$ changes more rapidly at lower $\eta_\text{in}/\eta_\text{out}$ than for the immiscible fluids. These observations for the miscible case are consistent with previous work~\cite{bischofberger2014fingering,Lajeunesse97,lajeunesse1999miscible} that claim added stability in the range $2/3< \eta_\text{in}/\eta_\text{out}<1$. This shift will be accounted for below by considering the structure of the inner fluid.

\section{Onset behavior}
This section explores what sets the length scale for the onset in each of the four cases described so far.

\subsection{Rectilinear immiscible}
\label{sec:linearonset}

\begin{figure*}
    \centering
    \includegraphics[width=.6\linewidth]{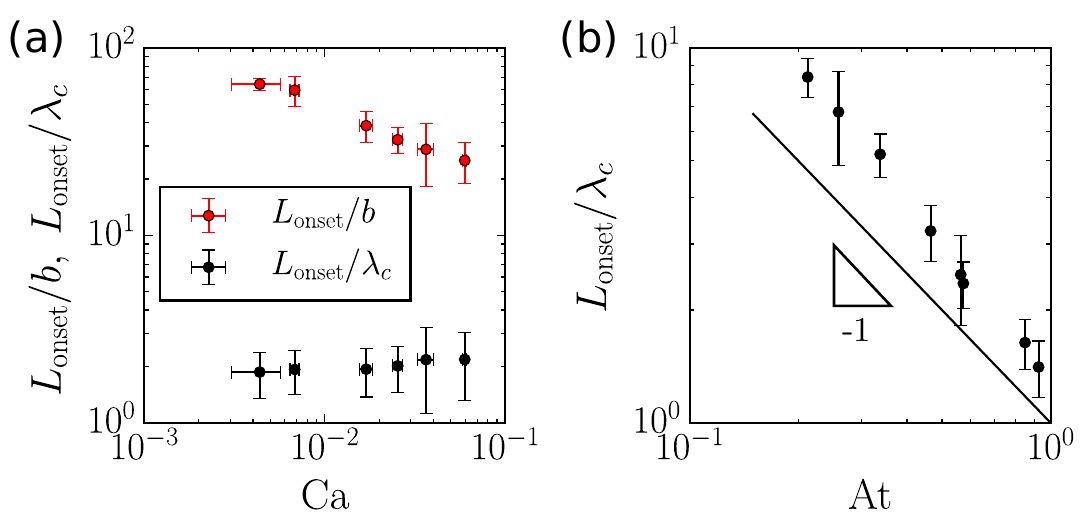}
    \caption{(a) shows the Ca dependence of $L_\text{onset}$ for the rectilinear immiscible system. Both $L_{\mathrm{onset}}/b$ (red) and $L_{\mathrm{onset}}/\lambda_c$ (black) are shown. (b) shows the Atwood dependence of $L_{\mathrm{onset}}/\lambda_c$, the black line has a slope of -1.}
    \label{fig:LinIM}
\end{figure*}

Above it was asserted that the correct length scale for the rectilinear geometry with immiscible fluids was \LonLc, instead of rescaling with the plate spacing, $b$. One can see this by looking at how the onset length at a fixed \vrat\ depends on the capillary number, defined as $\mathrm{Ca}=V\Delta\eta/\gamma$ where $V$ is the interfacial velocity and $\gamma$ is the interfacial tension.  This is shown in  Fig.~\ref{fig:LinIM}a. From this one can see that $L_{\mathrm{onset}}/b$ has the same dependence upon Ca as the most unstable wavelength, $\lambda_c$: $\lambda_c\sim \mathrm{Ca}^{-1/2}$. Dividing the onset length by the measured $\lambda_c$ it is seen that $L_\text{onset}/\lambda_c$ is a constant.

To gain understanding into the length scale for this onset point the classic linear-stability analysis of Saffman and Taylor~\cite{saffman1958penetration} is used. From this one can calculate $\Lambda_c$, the growth rate of the most unstable wavelength. Since perturbations grow exponentially, $e^{\Lambda_c t}$, then the timescale of the dynamics goes as $1/\Lambda_c$. If the interfacial velocity is constant then there is a characteristic length scale, $l_c = V/\Lambda_c$. The expression for this length scale is
\begin{equation}
    l_c = \frac{V}{\Lambda_c} = 3b\frac{\eta_{\mathrm{out}}+\eta_{\mathrm{in}}}{\eta_{\mathrm{out}}-\eta_{\mathrm{in}}} \mathrm{Ca}^{-1/2}.
\end{equation}
Noting that the most unstable wavelength is given by $\lambda_c=\pi b \mathrm{Ca}^{-1/2}$ and that the viscous Atwood number is defined as $\mathrm{At}=\frac{\eta_{\mathrm{out}}-\eta_{\mathrm{in}}}{\eta_{\mathrm{out}}+\eta_{\mathrm{in}}}$, I rewrite the critical length scale as
\begin{equation}
    \frac{l_c}{\lambda_c} = \frac{3}{\pi}\mathrm{At}^{-1}.
\end{equation}
Plotting our measurements for \LonLc\ versus At shows that the data is consistent with a power law of At$^{-1}$ with a prefactor close to unity. This argument says that the system is unstable as soon as the interface between the fluids is pushed. Moreover the length scale of when the growth saturates is set by the inverse growth rate from the linear-stability analysis. If one goes a step further and asks what the length scale should be when the perturbation grows to be about a wavelength in amplitude there is an additional $\log(\mathrm{Ca} \lambda_c /\epsilon_0)$ correction, where $\epsilon_0$ is some initial perturbation size. The choice of $\epsilon_0$ is a bit arbitrary, so this comparison is not included but the overall qualitative comparison provides indistinguishable scaling.

This comparison of $L_\text{onset}$, quantifying the transition between an exponential growth regime and linear growth, to a length scale from the linear stability analysis captures the dependence of both Ca and At that has been measured. However, this does not explain why the finger length needs to be around $0.3\lambda_c$ for linear stability to break down. I will examine if a similar explanation for the onset length can be seen in the miscible and radial systems.

\subsection{Rectilinear miscible}

At first glance there are two main differences between the rectilinear miscible and immiscible cases. The first is the wavelengths accessible to the two systems. For fluids with an interfacial tension $\lambda_c\propto\text{Ca}^{-1/2}$; for miscible systems $\lambda_c\sim 4b$ and is insensitive to injection and fluid parameters. The second is the added region of stability for miscible fluids when \vrat$\sim 2/3$. Given the arguments in the previous section, one might expect a similar magnitude of $L_\text{onset}/\lambda_c$, but with the miscible fluids having additional stability as \vrat$\rightarrow 1$. Comparing the magnitudes of the measured onset lengths, the miscible system becomes unstable at a length scale about three times larger than the immiscible system as well as having the expected shift with viscosity ratio, see Fig.~\ref{fig:OnsetData}b. Looking at the growth of fingers, as was shown in Fig.~\ref{fig:EarlyDelay}a and b, reveals another difference between these two cases: for immiscible fluids growth begins immediately, whereas for miscible fluids there is a delay length before the onset of dynamics.

\begin{figure*}
    \centering
    \includegraphics[width=\linewidth]{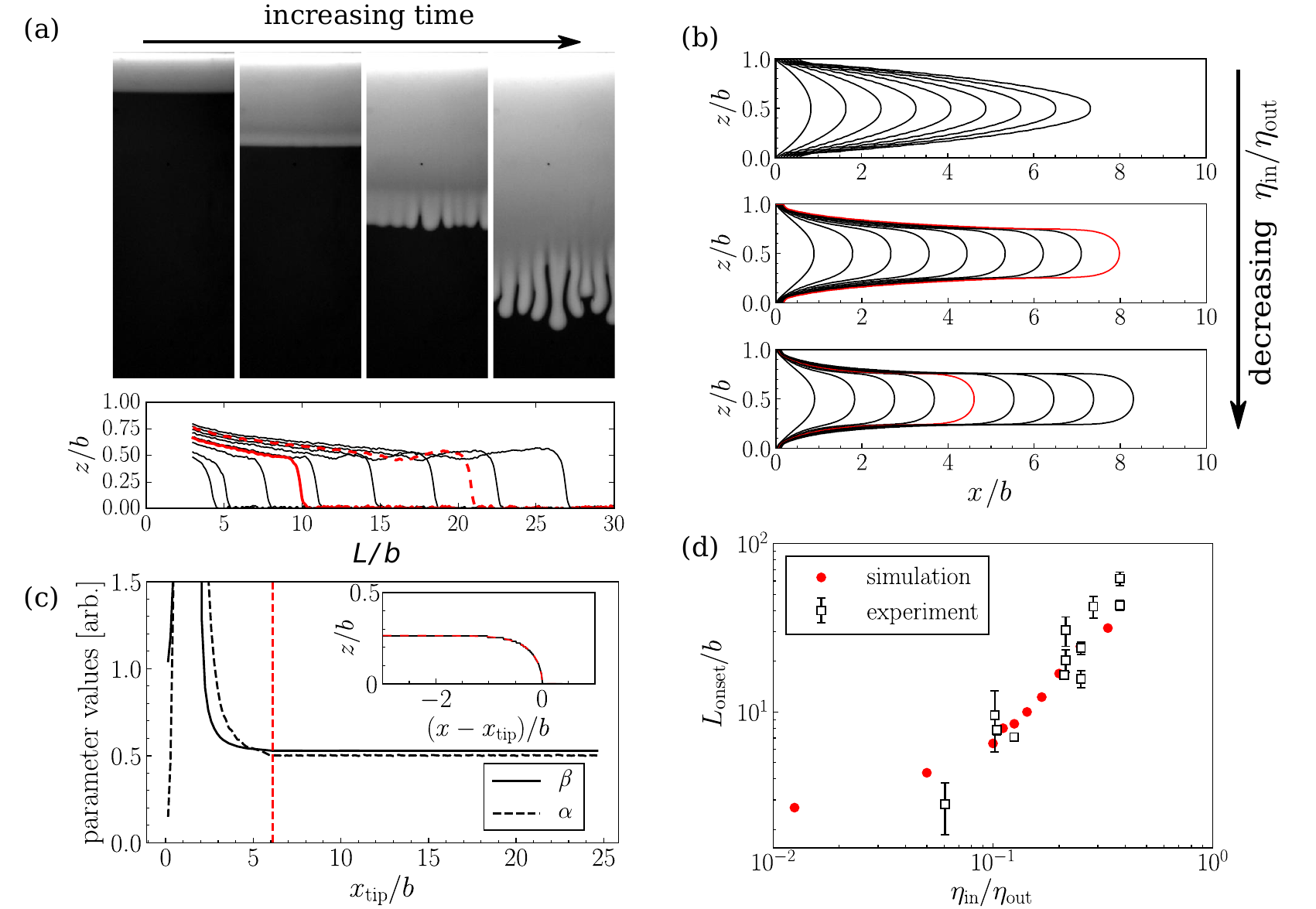}
    \caption{Panels in (a) show the evolution of an interface for miscible fluids in a rectilinear geometry. Below show the evolution of the thickness profile. The red line denotes the point where the bump at the tip first starts to form, the dotted red line is where $L_\text{onset}$ is measured from linear interpolation. In the images the lighter band at the interface in all but the first panel corresponds to this region of increased thickness. In (b) are results from COMSOL simulations showing the evolution of the inner fluid tongue for $\eta_{\mathrm{in}}/\eta_{\mathrm{out}}=1,\ 0.125,\ 0.05$. Each simulation is shown for the same amount of injection time. For the two lower viscosity ratios the tip of the inner fluid tongue reaches a steady state shape. (c) The change of the fit parameters $\beta$ and $\alpha$ as the profile evolves for $\eta_{\mathrm{in}}/\eta_{\mathrm{out}}=0.1$. After the vertical red dashed line the fit values become constant. The inset shows a comparison of the fit to the inner fluid tongue contour. (d) Shows a comparison between measured experimental onset lengths for miscible fluids based on the appearance of the bump at the tip and the onset of a steady state shape for the tip of the inner fluid in the simulations.}
    \label{fig:OnsetLinearMisc}
\end{figure*}

A place to look for a resolution to this increased onset length is in the dynamics of the inner fluid tongue structure. It has been seen in previous work~\cite{lajeunesse1999miscible,bischofberger2014fingering,videbaek2019diffusion} that the tip shape and the thickness of the inner fluid are important for the development of the fingering instability in the absence of surface tension. Experimentally it was found that a blunt tip of the inner fluid is always present whenever fingering occurs.  It has been speculated that this blunt tip is a necessary condition for the instability of the system. If this is the case, then the additional delay observed could be accounted for by the time needed for the system to develop this blunt structure.

The dynamics of the gap structure shown in Fig.~\ref{fig:OnsetLinearMisc}a shows that there is a period where the inner fluid tongue grows with a rounded tip until a blunt structure forms at the interface (the solid red line). At this point the tip grows thicker and fingers form. To see if these dynamics  occur independently of the lateral instability I numerically compute the flow within the gap. Similar dynamics are seen. The flow in the gap extrudes the inner fluid tongue as shown in Fig.~\ref{fig:OnsetLinearMisc}b. At a certain length scale from inception the tip acquires a steady state shape.

To determine when the simulations reach a steady state tip profile I look at how the tip shape evolves over time. It has been seen previously~\cite{rakotomalala1997miscible} that the shape of the interface is well matched by the following form~\cite{pitts1980penetration} in the steady state regime:
\begin{equation}
\exp\left( \frac{\pi x}{2 \alpha}\right)\cos\left( \frac{\pi z}{2\beta} \right) = 1
\end{equation}
where $x$ and $z$ are the coordinates of the interface, and $\alpha$ and $\beta$ are fitting parameters. I extract a contour of the interface and fit its positions within two plate spacings of the tip, $x-x_\text{tip} > -2b$ with $x_\text{tip}$ being the location of the interface at $z=b/2$. The inset of Fig.~\ref{fig:OnsetLinearMisc}c is a comparison of the fit form to the interface showing excellent agreement. Figure~\ref{fig:OnsetLinearMisc}c shows the evolution of $\alpha$ and $\beta$ as the inner structure develops. At a certain distance from the inlet both of these parameters reach a constant value.

Figure~\ref{fig:OnsetLinearMisc}d shows that the length scale when the simulation reaches a steady state matches well with when the experimental profiles first begin to thicken. In the simulations a subsequent widening of the tip is not observed; I believe that this additional structure is due to the onset of the fingering instability and that without including the lateral dimension this feature cannot be captured. In previous three-dimensional simulations for miscible fluids in a rectilinear geometry~\cite{Oliveira11} it was seen that there is tip thickening in the presence of fingering. From this I conclude that the miscible system first needs to develop a preferred interfacial shape within the gap before the lateral instability can begin. Only after this shape has been reached can fingers start to grow and the tip thickness begins to increase. The development of this structure occurs at a certain length scale and accounts for the delay in the fingering pattern.

\subsection{Radial geometry}
\label{sec:radialonset}

\textit{Immiscible fluids -} In the rectilinear geometry, immiscible fingering begins as soon as the interface begins to move while for miscible fluids there is an additional delay due to the necessity of forming structure within the gap. In the radial geometry, an additional delay is expected to occur since the fluid velocity is inversely proportional to the distance from the injection  point. This additional stabilization occurs because if a perturbation appears on the interface, then the peak will move more slowly than the trough, effectively adding a negative growth to the fingering process. In the linear-stability analysis this leads to a length scale at which the system should transition from being stable to being unstable. This length scale  will be compared to the data in Fig.~\ref{fig:OnsetData}a to see if it depends on \vrat, $b$, and $\lambda_c$ in the same way as does the experimental data. 

\begin{figure*}
    \centering
    \includegraphics[width=\linewidth]{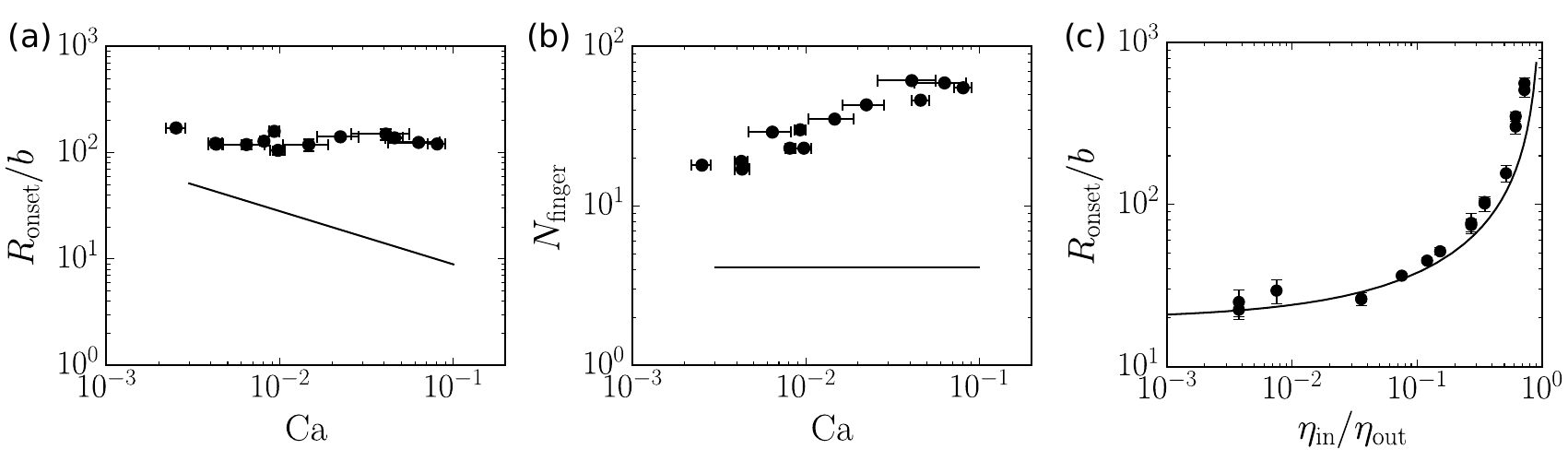}
    \caption{Onset in the radial geometry with immiscible fluids.  The data is for silicone oil invading a water-glycerol mixture with $\Delta\eta=828\ \mathrm{cP}$ and $\eta_{\mathrm{in}}/\eta_{\mathrm{out}}=0.38$. (a) The measured onset radius, $R_{\mathrm{onset}}/b$ versus Ca. (b) The number of fingers at onset,  $N_\text{fingers}$, is plotted versus Ca.  (c) $R_\text{onset}/b$ is plotted versus the viscosity ratio. The solid lines in each graph are the results of the linear stability analysis of Paterson~\cite{paterson1981radial}. The curve in (c) was calculated for Ca$=0.01$ and has been scaled by a factor of 4 to compare the functional form with the experimental data.}
    \label{fig:RadialIMOnset}
\end{figure*}

In order to compute the onset radius, one must start with a dispersion relation for the growth rate, $\Lambda$, that depends on the radius, $r$, and the mode number, $n$. Despite there being many linear-stability theories for radial viscous fingering~\cite{paterson1981radial,kim2009viscous,nagel2013new} calculations of the onset radius have not been conducted systematically. The onset radius is defined as the smallest radius at which any mode first acquires a positive growth rate. Noting that there is only one most unstable mode, $n_c$, this problem reduces to first finding this mode by computing $\partial_n\Lambda|_{n=n_c}=0$, and then finding the value of $r$ such that $\Lambda|_{n=n_c}=0$. The radius that is found corresponds to a transition from the system being stable to becoming unstable. For the following comparisons I use the theory of Paterson~\cite{paterson1981radial} but note that more recent theories that include additional boundary conditions or flows~\cite{kim2009viscous,nagel2013new} give quantitatively similar results and will not change the conclusions.

Figure~\ref{fig:RadialIMOnset} compares the experimental and theoretical results for the dependence of $R_\text{onset}$ on both Ca and \vrat.  Figure~\ref{fig:RadialIMOnset}a shows that the scaling and magnitude of the theoretical expectation for the Ca dependence is not consistent with our experiments. Appendix~\ref{sec:apptheory} shows that this conclusion is robust to the possible effects of noise in the data by calculating the effect of a finite-amplitude perturbation. Additionally Appendix~\ref{sec:appmeasure} shows how the transition length scale measured in Fig.~\ref{fig:OnsetData}a compares to the measurement of the stable-unstable transition shown in Fig.~\ref{fig:EarlyDelay}c. The transition length scale and the delayed onset exhibit the same scaling and only differ by a proportionality constant. 

A different check is to look at the number of fingers, $N_\text{fingers}$, that form in the experiment. This should be a reliable measure of the correct mode number for the instability. The results for $N_\text{fingers}$ shown in Fig.~\ref{fig:RadialIMOnset}b show the same discrepancy between theory and experiment; there is different scaling and different magnitude for the selected mode number between theory and experiment. 

Figure~\ref{fig:RadialIMOnset}c shows that there is better agreement between the Paterson theory~\cite{paterson1981radial} and experiment for how $R_{onset}/b$ depends on $\eta_\text{in}/\eta_\text{out}$. Although the functional form matches well, the theoretical curve has to be scaled by a constant value that depends on the Ca number used for the calculation.

\textit{Miscible fluids -} For the radial geometry, the values of $R_\text{onset}/b$ for the  miscible and immiscible cases show similar magnitudes for the onset, as shown in Fig.~\ref{fig:OnsetData}a, unlike the mismatch in magnitude seen for the rectilinear geometry, shown in Fig.~\ref{fig:OnsetData}b. The main difference between the two cases is the shift with viscosity ratio for the miscible case when the onset length begins to increase. Also, it appears that the dynamics of the inner fluid structure that are important for the linear miscible case are not necessary to account for the onset radius here; in the radial case the tip reaches a steady state well before fingering begins.

Though the behavior of $\lambda_c$ in the miscible limit had not been well accounted for by early work, recent theories that include viscous stress contributions to the pressure drop at the fluid interface~\cite{kim2009viscous,nagel2013new} are able to recover a saturation of $\lambda_c$ that is proportional to $b$ when $\text{Ca}>1$. However, these theories do not capture the shifted stability point at lower viscosity ratio and would not be able to model this feature of the experimental data.

\begin{figure*}
	\centering
	\includegraphics[width=\linewidth]{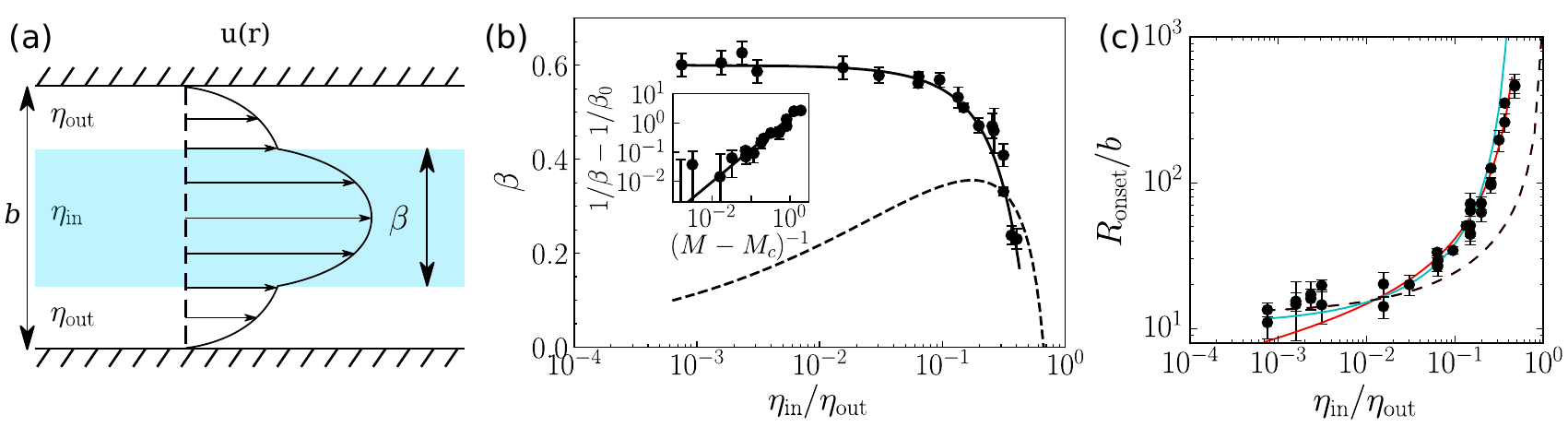}
	\caption{In (a) is a schematic of the flow profile within the gap for the calculation of the modified permeability. $\beta$ is the width of the inner fluid tongue. (b) Experimental measurements of the jump in viscosity contrast, $\beta$, for miscible fluids in a radial geometry over a range of $\eta_{\mathrm{in}}/\eta_{\mathrm{out}}$. The black line shows a best fit power law to a functional form of $(\beta^{-1}-\beta_0^{-1}) \propto (M-M_c)^{-\alpha}$ [$M=\eta_{\mathrm{out}}/\eta_{\mathrm{in}}$] where $1/M_c=0.51\pm0.08$, $\beta_0=0.60\pm0.01$, and $\alpha=1.1\pm0.3$; this form is seen in the inset. The dotted line is the prediction for $\beta$ given a kinematic-wave theory for the gap dynamics~\cite{lajeunesse1999miscible}. In (c) is the onset radius for miscible fluids compared to the linear-stability analysis of Paterson~\cite{paterson1981radial} using the modified permeability from Eqn.~\ref{eqn:DarcyPerm}. The blue line uses values of $\beta$ taken from experiment, in panel (b), and the red curves use $\beta$ derived from kinematic-wave theory~\cite{lajeunesse1999miscible}. The dotted line is from the Paterson linear-stability analysis without modifying the permeability.}
	\label{fig:RadialMiscOnset}
\end{figure*}

To account for this shifted stable point I add an additional ingredient by noting that for the miscible case the inner fluid tongue does not fully fill the gap between the plates~\cite{lajeunesse1999miscible,rakotomalala1997miscible,bischofberger2014fingering,videbaek2019diffusion,reinelt1985penetration}. The reason that this matters is that all quasi-two dimensional Hele-Shaw theories begin by averaging over Stokes flow in the gap to formulate Darcy's law.  This yields $\mathbf{u}=-\kappa\nabla P$ where $\kappa$ is the permeability derived from the gap averaging. In the limit of $\text{Ca}\ll 1$ for immiscible fluids the inner fluid will fill almost the entirety of the gap, except for a small wetting film, and results in a permeability of $\kappa = b^2/12\eta$. From Lajuenesse et al.~\cite{lajeunesse1999miscible} there are hints that only the presence of a blunt tip at the interface is relevant for the fingering instability to occur. I use the thickness of the tip to get a modified permeability. Assuming a piece-wise parabolic profile for the velocity field in the gap, Fig.~\ref{fig:RadialMiscOnset}a, similar to reference~\cite{lajeunesse1999miscible}, the permeability changes to
\begin{equation}
    \kappa_{i} = \frac{b^2}{12\eta_{i}}\left( \frac{\eta_{i}}{\eta_{\mathrm{out}}} + \beta^3 \left(\frac{\eta_{i}}{\eta_\text{in}}-\frac{\eta_{i}}{\eta_{\mathrm{out}}}\right)\right) = \frac{b^2}{12\eta_{i}\Omega}
    \label{eqn:DarcyPerm}
\end{equation}
where $\beta$ is the tip thickness, $\eta_i$ is the viscosity of the inner or outer fluid, and $\Omega$ is defined for brevity. Using this I modify the theory of Paterson and see how the onset radius depends on \vrat\ and compare that to experiment. As a note, for theories that include normal or tangential viscous stress at the interface~\cite{kim2009viscous,nagel2013new} the deviatoric component of the stress tensor would need to be multiplied by a factor of $\beta$ to account for a reduced width of the interface in the gap dimension.

In using this modified permeability the dependence of $\beta$ on \vrat\ needs to be known. The thickness, $\beta$, is measured a distance of $2b$ in back of the tip in the experimental thickness profiles once the patterns reach the onset radius. Measurements of $\beta$ for a range of \vrat\ are shown in Fig.~\ref{fig:RadialMiscOnset}b. As \vrat\ increases, the thickness at the tip tends toward zero. Fitting the data shows the drop to zero occurs at \vrat$=0.51\pm 0.08$, which is higher than previous experimental claims~\cite{bischofberger2014fingering} of 0.3 and closer to the theoretical prediction of 2/3. Precise measurements of this stability point become difficult as the onset radius diverges there. As \vrat$\rightarrow 0$ the thickness tends to a constant value of $\beta_{0}=0.60\pm0.01$, consistent with prior literature~\cite{taylor1961deposition,bretherton1961motion,rakotomalala1997miscible,reinelt1985penetration}. Another option for $\beta$ would be to use the value of a shock front height derived from kinematic-wave theory~\cite{lajeunesse1999miscible}, the dotted line in Fig.~\ref{fig:RadialMiscOnset}b.

With the adjusted permeability I compare the viscosity dependence of the onset to the modified theory, see Fig.~\ref{fig:RadialMiscOnset}c. It is seen that this change in the permeability provides a sufficient adjustment to the functional form to capture the effect of the shifted stability, though the magnitude from theoretical calculations is smaller and needs to be scaled by a constant value. In the limit of $\mathrm{Ca}\rightarrow\infty$ the modified theories of Kim et al.~\cite{kim2009viscous}, and Nagel and Gallaire~\cite{nagel2013new} are smaller by over an order of magnitude.

\section{Late time growth}

\begin{figure*}[b]
    \centering
    \includegraphics[width=.97\linewidth]{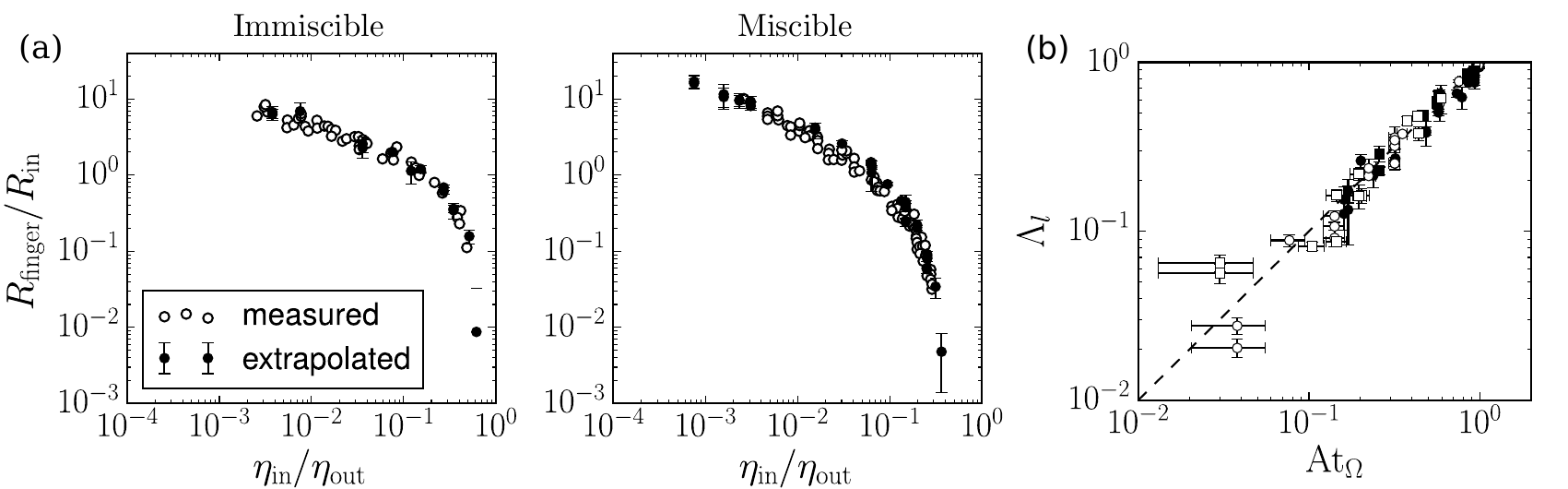}
    \caption{(a) Comparisons of the size ratio from previous experimental data (ref.~\cite{bischofberger2015island} for immiscible fluids (left) and ref.~\cite{bischofberger2014fingering} for miscible fluids (right)) (open circles) and the expected size ratio using our measured onset data and Eqn.~\ref{eqn:sizeratio} (black points). $R_{\mathrm{onset}}=315b$ is used in the calculation, corresponding to the reported pattern size. In (b) is the measured growth rates at onset plotted against At$_{\Omega}$. The dotted line has a slope of 1 and is a guide for the eye. The legend is the same as in Fig.~\ref{fig:OnsetData}.}
    \label{fig:RfRiComp}
\end{figure*}

In Section~\ref{sec:lineartransition} it was seen that the growth of fingers after the onset of the instability becomes linear with the pattern size. Noting this I make a simple model of the growth dynamics for the late-time regime. Up until the onset of the instability $R_{\mathrm{finger}}=0$ and after this point $R_{\mathrm{finger}}=\Lambda_l(R_{\mathrm{out}}-R_{\mathrm{onset}})$. Since previous work has looked at the size ratio, $R_{\mathrm{finger}}/R_{\mathrm{in}}$~\cite{bischofberger2015island,bischofberger2014fingering}, this model can be tested to see if it reproduces those results. Using the expressions above gives
\begin{equation}
\frac{R_{\mathrm{finger}}}{R_{\mathrm{in}}} = \frac{\Lambda_l(1-R_{\mathrm{onset}}/R_{\mathrm{out}})}{1-\Lambda_l(1-R_{\mathrm{onset}}/R_{\mathrm{out}})}.
\label{eqn:sizeratio}
\end{equation}
A similar construction can be done for the linear geometry. This expression is used in Fig.~\ref{fig:RfRiComp}a to compare the expected size ratio to those measured previously in a radial geometry for $R_{\mathrm{out}}=315b$. The quantitative values for the size ratio are captured well. Knowing the growth rates and the onset length scale allows us to predict the size ratio at any pattern size after the growth of fingering for both immiscible and miscible fluids.

In Eq.~\ref{eqn:sizeratio}, when $R_{\mathrm{out}} \gg R_{\mathrm{onset}}$ the size ratio becomes independent of the onset length. It was already seen in Fig.~\ref{fig:OnsetData}c that the growth rates for rectilinear and radial geometries look the same; the main difference arises from whether or not the system has a surface tension. From this one can infer that at late enough times in the pattern forming process the size ratio should be independent of geometry.

The difference in shifted stability between miscible and immiscible fluids can be reconciled by considering the thickness of the inner fluid tongue, as in Section~\ref{sec:radialonset}. Recall that in Section~\ref{sec:linearonset} the growth rate was proportional to the inverse of the viscous Atwood number. In the linear stability analysis the factors of $\eta$ that comprise the viscous Atwood number come from the permeability in Darcy's law. In Eq.~\ref{eqn:DarcyPerm} the effect of the tongue thickness essentially changed $\eta_{i}\rightarrow\Omega\eta_{i}$. Noting that $\Omega=1$ for the outer fluid, I construct a modified viscous Atwood number that encompasses the changing thickness of the inner fluid:
\begin{equation}
    \mathrm{At}_{\Omega} = \frac{\eta_{\mathrm{out}} - \Omega\eta_{\mathrm{in}}}{\eta_{\mathrm{out}} + \Omega\eta_{\mathrm{in}}}.
\end{equation}
Replotting the data from Fig.~\ref{fig:OnsetData}c against At$_{\Omega}$ shows a collapse of all the growth-rate data onto a single curve, Fig.~\ref{fig:RfRiComp}b. This single growth-rate curve provides a common link between all of the experimental systems including the size ratio that is observed at late times; all of these patterns behave in essentially the same way, as long as the inner fluid thickness is accounted for. One should keep in mind that this may change for miscible fluids once diffusion begins to become important since this has been seen to have quite dramatic effects on pattern morphology~\cite{videbaek2019diffusion}.
 
\section{Conclusion}

In this paper, the onset of the viscous fingering instability for both miscible and immiscible fluids was experimentally characterized in radial and rectilinear geometries. All of these cases show a commonality in the way that these patterns form; there is a length scale of the onset and transition to late time dynamics, and, once in this regime, fingers grow linearly with the size of the pattern. By combining these measurements, one can predict the size ratio of patterns well after the system has gone unstable. These quantities depend strongly upon the viscosity ratio, with miscible fluids exhibiting diverging behavior near \vrat$\sim 2/3$ rather than at unity as for immiscible fluids. This shift in stability is captured well by considering a modified permeability for Darcy's law that depends upon the thickness of the fluid tongue near the interface. The only case in which the onset length scale can be understood purely from existing theory is the rectilinear case with immiscible fluids.

A surprising finding of this work is in regards to the shortcomings of existing theoretical treatments when it comes to predicting the onset of fingering for the radial geometries. Despite the agreement that these works have shown with respect to capturing the wavelength of the instability, a common feature that is predicted is that for a given viscosity ratio, $R_{\mathrm{onset}}/b$ should decrease with increasing Ca. Instead, I have demonstrated that $R_{\mathrm{onset}}/b$ is constant, independent of Ca.

One place a possible resolution might be found is to look more carefully at the boundary conditions that contribute to the pressure drop at the interface. It has been seen that the inclusion of additional boundary terms has brought resolutions to incorrect scalings in the past~\citep{kim2009viscous,nagel2013new,dias2013wavelength}, specifically including additional normal and tangential stress balance at the interface. These changes helped to address the experimental observation that $\lambda_c$ becomes comparable to $b$ when $\text{Ca}\sim O(1)$ and is then independent of Ca. Even though these additional terms resolve the wavelength issue they are not adequate to describe the onset radius and finger number observed here. One route to pursue would be to consider the stress contributions that arise from the actual three-dimensional flow in the gap. With the rectilinear miscible system we have already seen that the flow in the gap was essential for determining the onset. For the immiscible case, the interface between the two fluids almost fully fills the gap. If the system is to satisfy no-slip at the top and bottom boundaries of the Hele-Shaw cell, then there must be a strong vertical component to the fluid velocity right at the interface, causing a rolling motion to advance it. This would locally cause higher shear at the interface resulting in larger stresses at the boundary which would be more than would be accounted for by only considering the normal stresses of the average flow. If there are indeed additional contributions from the gap dimension that are important for fully describing the observed patterns, then these features may not be applicable to other systems that exhibit branching growth.

Another important feature noted in this work is the significance of the structure at the tip of the inner fluid for the miscible case. In both radial and rectilinear geometries, the system acquires a particular structure before fingering begins. In the rectilinear case, the development of this structure is immediately followed by the instability, while in the radial case the system needs to reach a longer length scale before going unstable. Additionally we have seen that the shifted dependence on viscosity ratio for miscible fluids compared to immiscible fluids can be captured by accounting for the thickness of the inner fluid tongue and using a modified viscous Atwood number, $\mathrm{At}_{\Omega}$. Of particular interest is the fact that the onset radius and growth rate for miscible fluids seem to be so strongly tied to this thickness. This suggests that if one can control the shape of the tongue -- in particular make it thinner near its tip --  then it could be possible to lower the growth rate and increase the onset length.  This would provide additional control over the instability.

Control of instabilities, in particular turning them off, is a broad goal for many flow systems. By fully understanding what determines the onset of the instability, and subsequently pattern formation, one can begin to develop control schemes. This has worked exceedingly well in the case of immiscible fluids for the viscous fingering instability since surface tension has a well understood effect on the suppression of fingers~\cite{Li2009,Pihler2012,AlHousseiny12,Zheng2015,gao2019active}. However, miscible fluids have little to no options for stopping growth. In this paper we have highlighted how the tip structure at the interface of the viscous fingering pattern is coupled to the growth rate of fingers. This, and previous work on the importance of the gap structure~\cite{videbaek2019diffusion}, open up new possibilities for pursuing control in the case of miscible fluids in a high P\'eclet regime, where there is no apparent stabilization term.

\begin{acknowledgments}
I thank Irmgard Bischofberger, Rudro Rana Biswas, and Tom Witten for insightful discussions. I am greatly indebted to Sidney R. Nagel for his support, guidance, and illuminating conversations. The work  was supported by the University of Chicago Materials Research Science and Engineering Center, which is funded by the National Science Foundation under award number DMR-1420709 as well as the University of Chicago Physics Department through the Grainger Fellowship.

\end{acknowledgments}

\appendix

\section{Finite amplitude onset}
\label{sec:apptheory}

This section details how to account for a perturbation growing to a finite size in the theoretical framework. This is a less straight forward calculation since the growth rate for any given mode changes with radius. Typically the perturbation amplitude is defined as $\epsilon \exp(\Lambda(n,R) t)$, where $\epsilon$ is a small number, but due to a radial dependence of $\Lambda$ we define an amplitude, up to some $\epsilon$ as,
\begin{equation}
    A(n,r) = \exp{\int_0^t \Lambda(n,t')dt'} = \exp{\int_0^r\Lambda(n,r')r'dr'}
\end{equation}
where a constant areal injection rate is assumed. Here I use the growth rate from Paterson~\cite{paterson1981radial} due to its simpler form, though it should be noted that theories that contain additional contributions to the flow~\cite{kim2009viscous,nagel2013new} show qualitatively the same behavior. The non-dimensional growth rate equation has the following form
\begin{equation}
    \Lambda = \frac{-n(n^2-1)}{r^3}P\mathrm{At} + \frac{n}{r^2}\mathrm{At} - \frac{1}{r^2}
    \label{eqn:PatGrowth}
\end{equation}
where $n$ is the mode number (which we allow to be non-integer) and $P$ is a non-dimensional injection rate introduced in ref.~\cite{kim2009viscous} such that $P=1/(12\mathrm{Ca}r)$. The introduction of $P$ helps to account for the change in Ca as the pattern grows.

\begin{figure*}
    \centering
    \includegraphics[width=\linewidth]{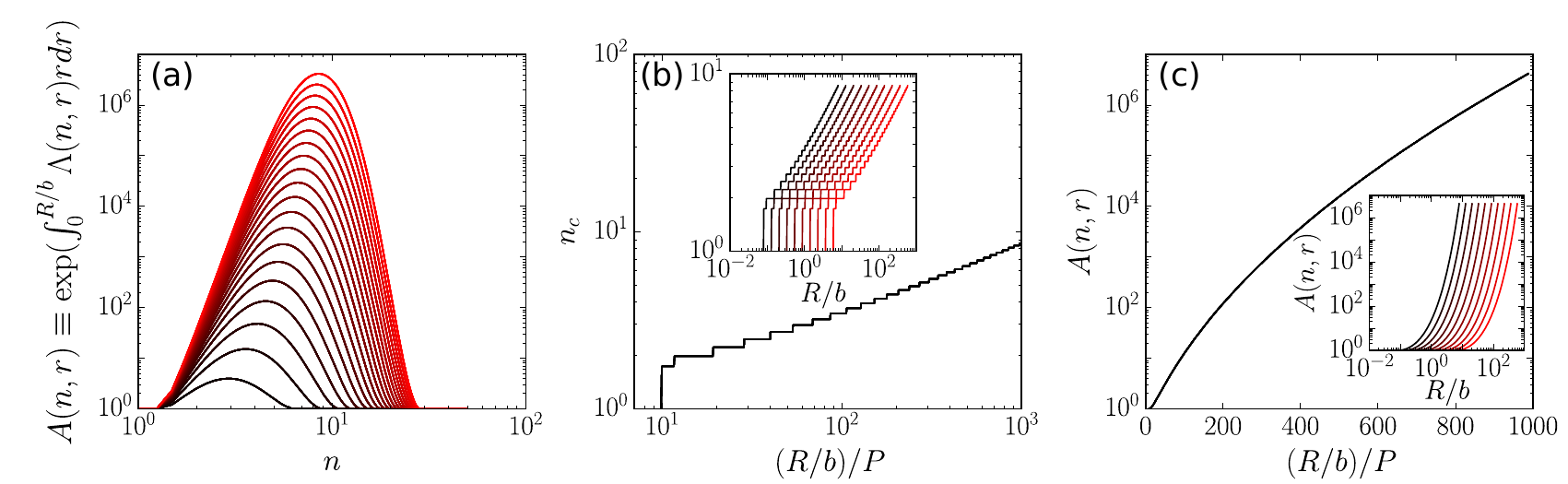}
    \caption{(a) is the accumulated amplitude in each mode, $n$, from whenever the first non-zero mode appears. This is done from radii of $R_{\mathrm{onset}}/b$ to $10R_{\mathrm{onset}}/b$. The transition of color from black to red corresponds to an increasing radius. (b) shows the mode, $n_c$, that has the largest accumulated amplitude as a function of radius. The start of all of these curves is the point where the interface first becomes unstable and scales with $P$. The inset shows the unscaled data. Lastly, (c) shows the maximum value of accumulated amplitude against the rescaled radius. The inset shows the unscaled data. The transition of color from black to red in the insets corresponds to increasing $P$.}
    \label{fig:FiniteAmp}
\end{figure*}

Using Eqn.~\ref{eqn:PatGrowth} one can see how the amplitudes for different modes change as a the pattern grows in size. In Fig.~\ref{fig:FiniteAmp}a the accumulated mode amplitude, $A(n,r)$, is shown at different radii. As $r$ increases the overall amplitude increases and the mode, $n_c$, that has the largest amplitude increases. The increase of $n_c$ is shown in Fig.~\ref{fig:FiniteAmp}b, note that even though the most unstable mode still increases, a change in $P$ merely shifts the onset radius but does not change the behavior of $n_c$ after this point. This is similarly seen in Fig.~\ref{fig:FiniteAmp}c where the maximum amplitude is shown as a function of $r$. Changing $P$ changes the onset radius, but once fingers form they all grow in a similar manner. This is inconsistent with experimental data shown in Fig.~\ref{fig:RadialIMOnset} where a strong dependence on Ca is seen for $N_\text{fingers}$ as well as the observation that the onset radius is independent of Ca. This means that even accounting for measurement of a finite amplitude, at least in the context of the linearly unstable evolution of the fingering instability, does not account for the discrepancies between theory and experiment.

\section{Stability-instability transition point}
\label{sec:appmeasure}

In the text, a distinction was drawn between two length scales, one being the transition to late time growth and the other being a delayed onset associated with a stable-unstable transition. The latter is characterized by the system being stable for radii smaller than the onset radius and the system being unstable to fingering afterwards. In Section~\ref{sec:radialonset} comparisons of the transition length were made to theoretical predictions of the delayed onset. Here are described two measurements that can be made of the delayed onset and how these compare with the measurements of the transition length shown in Fig.~\ref{fig:OnsetData}a.

The first is a threshold measurement. In the methods section the description of $R_\text{finger}$ mentioned that the initial conditions were subtracted off and that only growth from this initial state is shown. When looking at the subsequent growth there is a finite length scale when the system grows past its initial conditions, as was seen with the rectilinear miscible data in Fig.~\ref{fig:EarlyDelay}b. This same type of delay is seen in the radial case.

Another method, described in Section~\ref{sec:lineartransition} and shown in Fig.~\ref{fig:EarlyDelay}c, uses the expected value of the most unstable wavelength, $\lambda_c$. In previous literature it has been seen that the wavelength in both rectilinear and radial geometries match well with theoretical predictions. In the radial case we can use a geometrical constraint to measure when the fingers should have first appeared. Looking at a single finger we can measure its angular extent and its average radial position. With these two quantities we can infer the arc length of that segment, which we will call $\lambda_s$. For the immiscible case the interfacial velocity is measured, allowing one to get a local value of Ca, and from this calculate the expected wavelength. In Fig.~\ref{fig:EarlyDelay}c was shown the evolution of $\lambda_s$ and $\lambda_c$ as a function of the pattern size. Those curves cross at a single radius, which should be the only allowed point at which the finger could have started growing.

\begin{figure*}
    \centering
    \includegraphics[width=.4\linewidth]{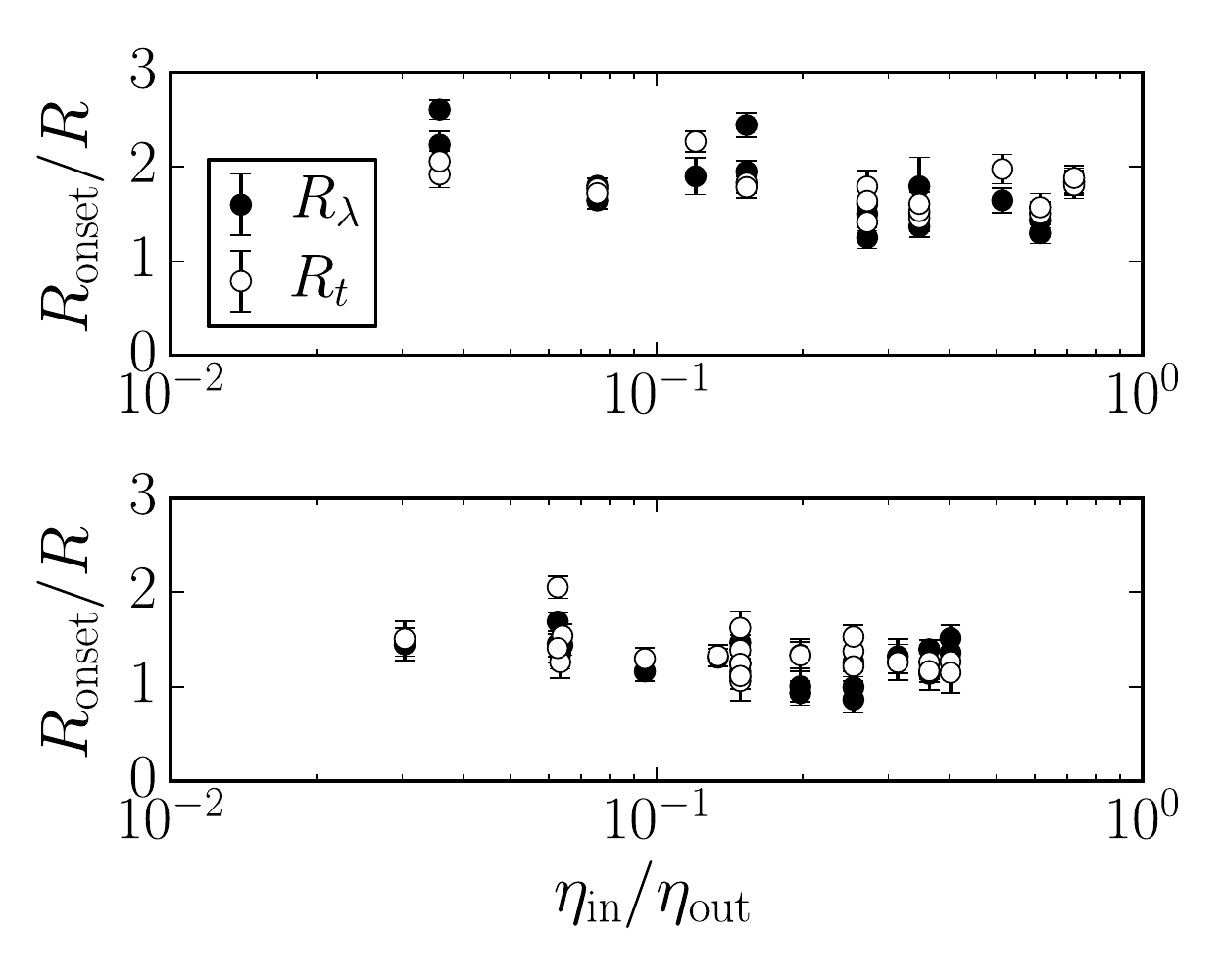}
    \caption{
    This figure compares the value of the onset length for the radial geometry measured by linear interpolation to the delay length measured by looking at the finger size, $R_t$, (the white points) and from the $\lambda$ crossover measurement, $R_\lambda$, (the black points). This is done for both immiscible (top) and miscible (bottom) systems.}
    \label{fig:OnsetCompApp}
\end{figure*}

With these two methods of measurement we can compare them to each other as well as the linear interpolation method for both immiscible and miscible fluids. Fig.~\ref{fig:OnsetCompApp} shows the ratio of $R_{\mathrm{onset}}$ (linear growth onset) to $R_t$ (threshold) and $R_{\lambda}$ (wavelength matching). The ratio of $R_{\mathrm{onset}}/R_{t}$ and $R_{\mathrm{onset}}/R_{\lambda}$ are consistent with each other, meaning that theses two additional measurement methods are measuring the same point up to within noise. We also see that the ratio is independent of \vrat, meaning that the transition length scale differs from the delayed onset by only a constant factor.

\bibliography{main.bib}

\end{document}